\documentclass[12pt,preprint]{aastex}
\slugcomment{Not to appear in Nonlearned J., 45.}

\shorttitle{Dust attenuation in the nearby universe}
\shortauthors{Buat et al.}

\begin{document}

\title{Dust attenuation in the nearby universe: \\ 
comparison between galaxies selected
in the ultraviolet 
or 
in the far-infrared}


\author{V. Buat\altaffilmark{1}, J. Iglesias-P\'{a}ramo\altaffilmark{1}, M.
Seibert\altaffilmark{2}, D. Burgarella\altaffilmark{1},S.
Charlot\altaffilmark{3,4}, C.
Martin\altaffilmark{2},C.K. Xu\altaffilmark {2}, T. M. Heckman\altaffilmark{5},
S.
Boissier\altaffilmark{6}, 
A. Boselli\altaffilmark{1}
T. Barlow\altaffilmark{2}, L. Bianchi \altaffilmark{5}, Y.-I. Byun\altaffilmark
{7}, J. Donas\altaffilmark{1}, K. Forster\altaffilmark{2}, P. G. 
Friedman\altaffilmark{2}, P. Jelinski\altaffilmark{8}, Y. -W.
Lee\altaffilmark{7}, B. F. Madore\altaffilmark{6}, R. Malina
\altaffilmark{1}, B.
Milliard\altaffilmark{1}, P. Morissey\altaffilmark{2}, S.
Neff\altaffilmark{9}, M. Rich\altaffilmark{10}, D.
Schiminovitch\altaffilmark{2}, O. Siegmund\altaffilmark{8}, T.
Small\altaffilmark{2}, A. S. Szalay\altaffilmark{5}, B. Welsh\altaffilmark{8}
and
T. K. Wyder\altaffilmark{2}}

\altaffiltext{1}{Laboratoire d'Astrophysique de Marseille, Marseille 13012,
France
(veronique.buat,jorge.iglesias,denis.burgarella,alessandro.boselli,jose.donas,roger.malina,
bruno.milliard@oamp.fr)}
\altaffiltext{2}{California Institute of Technology, MC 405-47, 1200 East
California Boulevard, Pasadena, CA 91125
(mseibert,cmartin,tab,krl,friedman,patrick,ds,tas,wyder@srl.caltech.edu,cxu@ipac.caltech.edu)}
\altaffiltext{3}{Institut d'Astrophysique de Paris, 75014 France}
\altaffiltext{4}{Max Planck Institut f\"{u}r Astrophysics, D-85748 Garching,
Germany (Charlot@MPA-Garching-MPG.DE)}
\altaffiltext{5}{Department of Physics and Astronomy, The Johns Hopkins
University, Homewood Campus, Baltimore, MD 21218
(heckman,bianchi,szalay@pha.jhu.edu)}
\altaffiltext{6}{Observatories of the Carnegie Institution of Washington,
813 Santa Barbara St., Pasadena, CA 9110
(boissier@ociw.edu,barry@ipac.caltech.edu)}
\altaffiltext{7}{Center for Space Astrophysics, Yonsei University, Seoul
120-749, Korea (byun,ywlee@obs.yonsei.ac.kr)}
\altaffiltext{8}{Space Sciences Laboratory, University of California at
Berkeley, Berkeley, CA 94720(patj,bwelsh,ossy@ssl.berkeley.edu)}
\altaffiltext{9}{Laboratory for Astronomy and Solar Physics, NASA Goddard
Space Flight Center, Greenbelt, MD 20771 (neff@stars.gsfc.nasa.gov)}
\altaffiltext{10}{Department of Physics and Astronomy, University of
California,
Los Angeles, CA 90095 (rmr@astro.ucla.edu)}

\begin{abstract}
We compare  the dust attenuation properties of   two
samples of galaxies purely
selected in the near-ultraviolet (NUV)  band (1750-2750 $\rm \AA, 
  \lambda_{mean} =2310   \AA$) and in the  far-infrared
(FIR)  at 60 $\mu$m . These samples are built
using the 
GALEX and IRAS 
sky surveys over $\sim$ 600 square degrees. The NUV selected 
sample contains
95  galaxies detected down to $\rm NUV=16$ mag (AB system).  83 
galaxies in this sample are spirals or irregulars and 
 only two of them are not  detected at 60 $\mu$m. The FIR selected sample is
 built from the IRAS PSCz catalog complete down to 0.6 Jy. Among the 163
sources, 
we select 118 star forming galaxies well measured by IRAS, all but 1 are
detected in NUV and  14 galaxies are not detected in the far-ultraviolet (FUV) 
band (1350-1750 $\rm \AA, \lambda_{mean} =1530  \AA$). 
The dust to ultraviolet (NUV and FUV) flux ratio 
is calibrated to
estimate the dust attenuation at both wavelengths.
The median value of the  attenuation  in NUV is found to be
$\sim 1$ mag for the NUV selected sample versus $\sim 2$ mag for the FIR
selected one.
Within both samples, the dust attenuation is found to correlate with the 
luminosity  of the galaxies. Almost all the 
NUV selected  galaxies and  2/3 of the FIR selected sample exhibit a lower dust
attenuation than that expected from
the tight
relation found previously for starburst galaxies 
 between the dust attenuation and the slope of the ultraviolet continuum. The
situation is  inverse for one third of the  FIR selected galaxies: their
extinction is higher than that deduced from  their FUV-NUV color and the
relation valid for starbursts.
\end{abstract}


\section{Introduction}

To quantify the star formation activity in the Universe, from low to high
redshift or in individual galaxies we need an estimate as accurate as possible
of the recent star formation rate in galaxies.
Among  various potential estimators, the far-infrared and ultraviolet
luminosities are commonly used. Both emissions are expected to come from young
stars (e.g. Kennicutt 1998 and references therein). The ultraviolet light is
emitted by young stars and is theoretically
directly
connected to the recent star formation rate.  
However, the use of the ultraviolet emission to trace the star formation
is largely hampered by the presence of the dust which absorbs and scatters the
UV light; the far-infrared emission is not
affected by this limitation. In fact the ultraviolet  and FIR emissions are
complementary: the ultraviolet  light lost because of the dust absorption is
re-emitted in
FIR by the dust. This energy budget has been used to derive a measure of the
dust  attenuation \citep{xub95,meu99,gor00}. This attenuation is known to be
very large in some
objects and even if  it remains modest in average in the nearby universe
\citep{xub95} 
it is crucial to correct the observed ultraviolet  flux before any quantitative
interpretation. 

 In this
paper we  consider a galaxy sample  selected in the NUV for which most of the
galaxies have been observed in FIR. Because of its selection, this sample will
be directly
comparable to optically (ultraviolet rest-frame)  selected
samples at higher redshift.  We will 
derive the
dust attenuation  from the comparison between  the far-infrared and the
ultraviolet   
emissions and discuss  variations of this attenuation with  luminosity and
FUV-NUV color of the galaxies.
For comparison, we will also build a FIR selected sample of galaxies, most of
these galaxies  being  detected in   NUV.  We will
perform the
same analysis as for the NUV selected sample

A similar pioneering work was already performed over only 35 deg$^2$ and in a
single
ultraviolet wavelength \citep{igl04}. Complementary approachs may consist in  
exploring variations of the FIR/UV ratio within single 
nearby galaxies (Boissier et al., Popescu et al. this volume)

\section{The samples}

Here, we briefly discuss how we select our samples. 
Galaxy distances are calculated from their redshift measurements taken from
the NASA/IPAC Extragalactic Database
(NED), HyperLeda (http://leda.univ-lyon1.fr)   and the IRAS PSCz \citep{sau00}
and assuming $H_0 = 72~ {\rm km s^{-1} Mpc^{-1}}$. For very
 nearby galaxies (4 sources)  their distances were found in the literature. 
 All magnitudes are expressed in the AB system.\\

%\subsection{The NUV selected sample}

Our NUV selected sample, down to $\rm  NUV = 16$~mag,  is built from 
 the first observations  of the GALEX  All Imaging Survey (AIS). See Martin et
al. and Morissey et al. (this volume) for details regarding the GALEX
instrument and mission. We select only
the frames
with
exposure
times larger than or equal to 50~sec: the resulting field
covers  615 deg$^2$. We perform photometry using the IR0.2 calibration and 
apply a zero point offset of +0.039 mag in FUV and -0.099 mag in NUV.  We mask
bright
stars and
galaxies lying close to our  targets. For
each target, the sky background is measured by combining
 several individual nearby sky regions. NUV fluxes are extracted using
elliptical apertures enclosing the total fluxes.
FUV magnitudes are measured
within the same aperture. 

NUV and FUV fluxes are  first corrected for
Galactic extinction
 using  the Schlegel et al. (1998) dust map and the Cardelli et al. (1989)
extinction
curve. A total of 95
sources brighter than $\rm NUV = 16$~mag  are then selected. We exclude objects
whose ultraviolet  flux
can be
contaminated by other sources than young stars i.e. early type (2 ellipticals)
and active galaxies (2 Seyfert~1, 2  Seyfert~2 and 1 QSO). Most of the
remaining
88
sources are late-type spirals.  

We search for a detection at 60 $\mu$m for all the 88 sources using the IRAS
FSC \citep{mos90} and the  Scan Processing and Integration Facility (SCANPI). 4
galaxies were not observed by IRAS. All other
sources  but 3  are detected at 60 $\mu$m. 
For the 3  non detections, we adopt
a conservative upper limit of 0.2 Jy at 60$\mu$m. 
Among the 81 galaxies with an IRAS detection, 22 are contaminated by cirrus
or have close neighbours not resolved by IRAS.
 We discard these measures from the quantitative
analysis of the FIR emission. 
We are thus left with 59 sources with a NUV and a  flux at 60 $\mu$m and 3 NUV
sources 
with only an upper limit at 60 $\mu$m. \\

 %\subsection{The FIR selected sample}

To construct the FIR selected sample, we
start
with the IRAS
PSCz  catalog \citep{sau00} which is complete down to 0.6 Jy at 60 $\mu$m; we  
take advantage of the optical
identification of the
IRAS sources  to search for a NUV detection. We
select 163 IRAS sources over 509 deg$^2$  detected at 60
$\mu$m with a
reliability  larger than 50\%  and not contaminated by   cirrus. 144  targets
also have
a flux at 100 $\mu$m and the
remaining have only upper limits. Among these 163  galaxies, 97 have a
morphological type in NED or LEDA.
We discard  4  ellipticals or lenticulars,
3 Seyfert 1 and 5  Seyfert 2 and  33 galaxies which have close neighbours not
resolved by IRAS. 
We are thus left with 118 galaxies. 
 NUV and FUV photometries  are performed following the same prescriptions as
for the NUV
selected sample.

All
but 1 galaxy  are detected in NUV (S/N ratio larger than 3). 14
galaxies are not detected in FUV.
 For the non detections in NUV and/or FUV  upper
limits are estimated corresponding to a S/N ratio of 3.

\section{The dust attenuation  at ultraviolet  wavelengths}
\subsection{Modeling}
It  was shown previously 
\citep{bua96,meu99,gor00}  
 that for ultraviolet  measurements up to 2000 $\rm \AA$ the 
dust to ultraviolet flux  ratio is a robust tracer of  dust attenuation in star
forming 
galaxies regardless of 
the details of the extinction mechanisms (dust/star geometry, dust
properties).  Hereafter $F_{\rm dust}$   refers to the total dust emission
and 
$F_{\rm FUV}$ and $F_{\rm NUV}$
are defined for each GALEX band as $\nu F_{\nu}$ where $F_\nu$ is expressed in
$\rm W m^{-2} Hz^{-1}$. 

 Since the NUV band of GALEX lies at longer wavelengths than 2000 $\rm \AA$, 
we must
test the
reliability of the $F_{\rm dust}/F_{\rm NUV}$ ratio to estimate the dust
attenuation for 
various scenarii of star formation. 
At this aim, we use the population synthesis  code PEGASE \citep{fio97} under
 different 
hytotheses of star formation rate (SFR) (Figure 1).
 The amount of dust emission is
obtained by adding all the stellar emission absorbed by the dust. We try 
different  configurations for the dust attenuation: the  screen-like Calzetti's
attenuation law \citep{cal00},  a homogeneous mixture of dust and stars in a
slab geometry
with a Galactic or a LMC stellar extinction law \citep{bua96} or a clumpy
medium \citep{cal94}. We also include 
 the time dependent scenario  proposed by Charlot \& Fall (2000).
Within this frame, we develop another bi-variate model
in which the attenuation  of the light from  stars younger than $10^7$ years
follows the
Calzetti's law whereas
 older stars are supposed to be homogeneously mixed with the dust  or to be
distributed in a clumpy medium. All these models give very
similar results for a given star formation history with a dispersion lower than
5 $\%$. In particular the exact form of the extinction curve (with or without
a 2175 $\rm \AA$ bump) does not affect the results. Each curve plotted in
Figure 1 is the average over all these scenarii for a specific star formation
rate. 
Similar plots are found with the FUV
flux instead of the NUV one. Only the exponential decrease of the
SFR with $\tau$= 2 Gyr leads to a much lower attenuation for a given dust to
NUV 
flux ratio: the dust heating by old stars becomes important and only a
fraction of the dust emission is related to the ultraviolet  absorption. Such a
steep
decrease of the SFR is typical of  elliptical or lenticular galaxies
\citep{gav02}.
All the other scenarii of star formation lead to a relative error lower  than 
20\%. The calibration
holds for both GALEX bands and the  $F_{\rm dust}/F_{\rm NUV}$ will be used to
estimate the dust attenuation since more galaxies are detected in NUV than in
FUV. 

We perform polynomial fits  on our models excluding the curves
with $\tau =$2 Gyr since the FIR and the NUV selections focus on active star
forming
galaxies and we exclude ellipticals and lenticulars from the study. 
\begin{equation}
A({\rm NUV}) = -0.0495x^3+0.4718x^2+0.8998x+0.2269
\end{equation}
where $x=\log(F_{\rm dust}/F_{\rm NUV})$ and 
\begin{equation}
A({\rm FUV}) = -0.0333y^3+0.3522y^2+1.1960y+0.4967
\end{equation}
where $y=\log(F_{\rm dust}/F_{\rm FUV})$.
These fits are  fully consistent with the mean relations between the dust
 attenuation and the dust to ultraviolet flux ratio for both GALEX bands
proposed by Kong et al. (2004) and Charlot \& Brinchmann (private
communication) for similar star formation histories.

\subsection{ The amount of dust attenuation }

Before estimating the attenuations with the above formulae, we need to estimate
the total dust emission. For the
galaxies observed at  60 and 100 $\mu$m we calculate this total dust emission
following
Dale et al. (2001). 3 
galaxies from the NUV selected sample were not detected at 100 $\mu$m: we use
the
mean value of the $F_{60}/F_{100}$ ratio found for the sample galaxies
detected at both wavelengths  to estimate their flux at
100 $\mu$.  The mean
bolometric corrections 
$<F_{\rm dust}/F_{\rm FIR}> $  are equal to $2.8\pm 1.1$ and $2.4\pm 1.1$  for
the NUV and FIR selected sample respectively (the FIR flux is defined between
40 and
120 $\mu$m \citep{hel88}). These values are  intermediate between the
values found for starbursts and cirrus dominated galaxies \citep{cal00,row03}.

In the NUV selected sample, 
the median values of the dust attenuation in GALEX bands are found to be 
$0.8^{-0.3}_{+0.3}$  mag in NUV and  and $1.1^{-0.4}_{+0.5}$ mag in FUV. 

In the FIR selected sample, the
median values of the  dust attenuation are found larger:  $2.1^{-0.9}_{+1.1}$
mag in NUV and
$2.9^{-1.1}_{+1.3}$ mag in FUV.
The  galaxies not detected in FUV have a 
dust attenuation larger
than $\sim 5$  mag in agreement with the large attenuations  ($\ge\sim$4 mag)
found in NUV. The
histograms of the attenuations are given in Figure 2.

In Figure 3, the dust attenuation in the NUV band is compared to the intrinsic
luminosity of the galaxies expected to be the best measurement of  the recent
star formation rate: the NUV luminosity corrected for the extinction for the
NUV selected sample (left panel) and the dust luminosity for the FIR selected
sample (right panel). In both cases the dust attenuation is found to increase
with
the   luminosity.   This result is 
consistent with  findings of an increase of the dust
  attenuation with the star formation rate in nearby galaxies
\citep{wan96,hop01}. Vijh et al. (2003) proposed a law between the extinction
and the
 luminosity at 1600 $\AA$ for bright Lyman break galaxies at $z>2$: the 
extrapolation
 of this law to fainter objects predicts
 lower attenuations than  those found for our NUV selected sample. This effect
might be attributed to some evolution of the dust attenuation with redshift. 
 
\subsection{The mean dust attenuation in the local Universe}

We can compare the dust attenuation found in our samples to the mean dust
attenuation estimated from the FIR and NUV (or FUV) luminosity densities (e.g.
Buat et al. 1999). At this aim we calculate the ratio of the FIR and NUV (FUV)
luminosity densities and translate it to a mean dust attenuation. The FIR
density is taken from Saunders et al. (1990): $\rm \rho_{FIR} = 3.9 ~10^7 
L\sun /Mpc^3$ with h=0.7. The NUV and FUV luminosity density  are derived from
the new GALEX data (Wyder et al., this volume): $\rm \rho_{NUV} = 1.8  ~10^7 
L\sun /Mpc^3$  and $\rm \rho_{FUV} = 1.9 ~10^7 L\sun /Mpc^3$. Adopting $<F_{\rm
dust}/F_{\rm FIR}>\simeq 2.5$ we obtain $\rm \rho_{dust}/\rho_{NUV}\simeq
\rho_{dust}/\rho_{FUV}\simeq 5.5$ which translates to dust attenuations of
 1.1 mag and 1.6 mag in NUV and FUV respectively. These mean values are
intermediate between those found with our  FIR and NUV selection although
closer to the dust attenuation found for the NUV selected galaxies. Therefore, 
as far as the dust attenuation is concerned, a NUV selection seems to be more
representative of the mean galaxy properties in the local universe than a FIR
selection. 
\subsection{ FUV-NUV color and dust attenuation}

Bell (2002) and more recently Kong et al. (2004) showed that the slope of the
UV continuum (which can be traced by the FUV-NUV color) is not a reliable
tracer of the dust  attenuation in galaxies which
are not experiencing a strong starburst.
In Figure 4  is plotted the FUV-NUV color   against the dust to FUV flux ratio
for the NUV selected and the FIR selected galaxies. We  keep only galaxies
brighter than 20 mag in NUV and FUV in order to have a photometric  error in
each band  lower than  or equal to 0.15
mag. 
 Almost all the  NUV selected galaxies and a large fraction of the FIR selected
ones lie below the curve followed by   starburst
galaxies \citep{meu99,kon04}: they  exhibit  a lower $F_{\rm dust}/F_{\rm FUV}$
(i.e. a
lower  attenuation) than starbursts for a given
FUV-NUV color, as already found for non starbursting galaxies. Kong et al.
(2004) explain this behavior by a variation of the ratio of present to past
average star formation rate: our data are roughly consistent with a ratio
varying from 1 to 0.1. 
 About 1/3 of the  FIR selected galaxies
lie above the curve valid for starburst galaxies: their dust attenuation is
larger
than that expected from the FUV-NUV color. This trend is similar to that  for
Ultra-Luminous Infrared Galaxies \citep{gol02} although the dust attenuations
found
here are not as large as for the ULIRGs. The FUV-NUV color seems therefore
unable
to trace
the  dust attenuation in these galaxies.

\acknowledgments
GALEX (Galaxy Evolution Explorer) is a NASA Small Explorer, launched in April
2003.
We gratefully acknowledge NASA's support for construction, operation,
and science analysis for the GALEX mission,
developed in cooperation with the Centre National d'Etudes Spatiales
of France and the Korean Ministry of 
Science and Technology.

%\clearpage

 \begin{figure}
%\epsscale{.80}
\includegraphics[angle=-90]{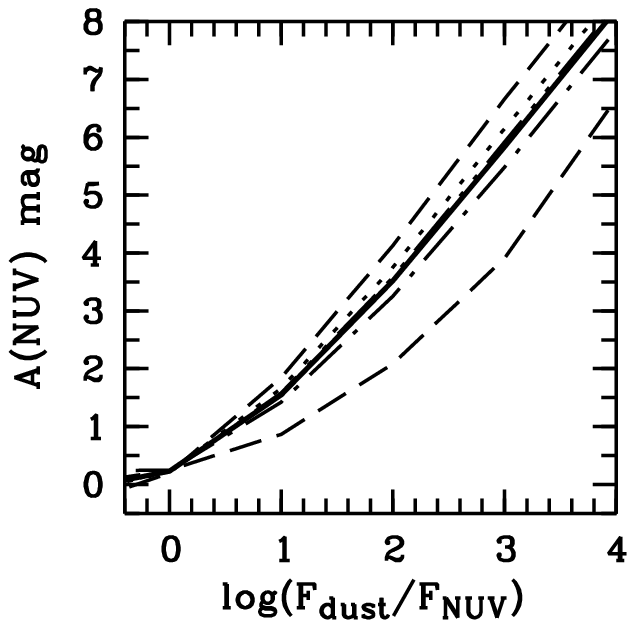}
   \caption{The predicted attenuation by the dust  in the GALEX NUV band as a
function of the
dust to  NUV flux ratio.  Various star formation rates are
 used: constant (dotted line), a 5 Myr old burst 
(upper short-dashed  line), exponential decrease with $\tau=$2 Gyr (lower short
dashed line), 4
Gyr (dot-dashed line)  and 8 Gyr (solid line).  The
polynomial fit adopted in the paper (eq.(1)) is plotted with a solid line and
closely follows the $\tau=$8 Gyr model}
      \label{fig1}
   \end{figure}

 \begin{figure}
\includegraphics[angle=-90]{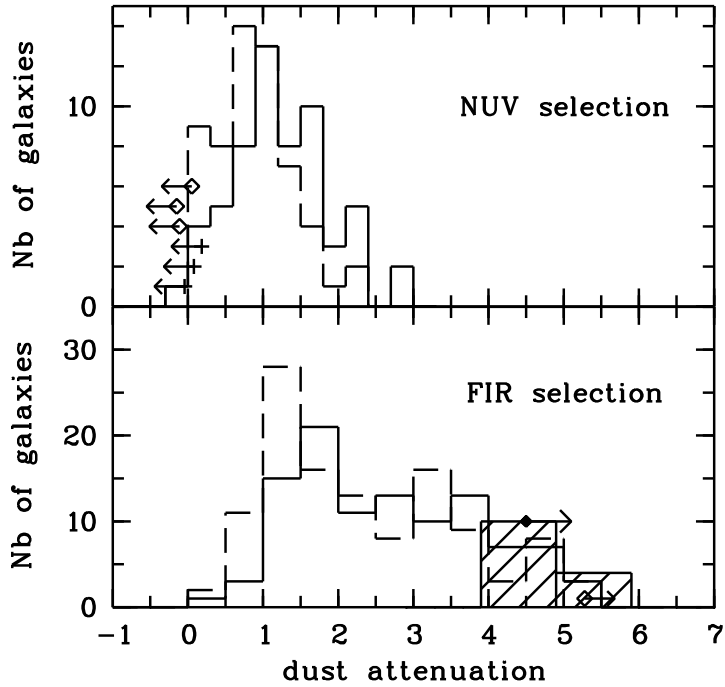}
   \caption{ Histograms of the dust attenuation. Solid line: A(FUV), dashed
line: A(NUV). Upper panel:  NUV selected sample, the upper limits are
plotted as crosses and left arrows for the FUV band and diamonds and left
arrows for the NUV band. Bottom panel: FIR selected sample,  the dashed
histogram represents the lower limits in A(FUV), the only lower limit in
A(NUV) is plotted as a diamond and a right arrow}
      \label{fig2}
   \end{figure}
 \begin{figure}
\includegraphics[angle=-90]{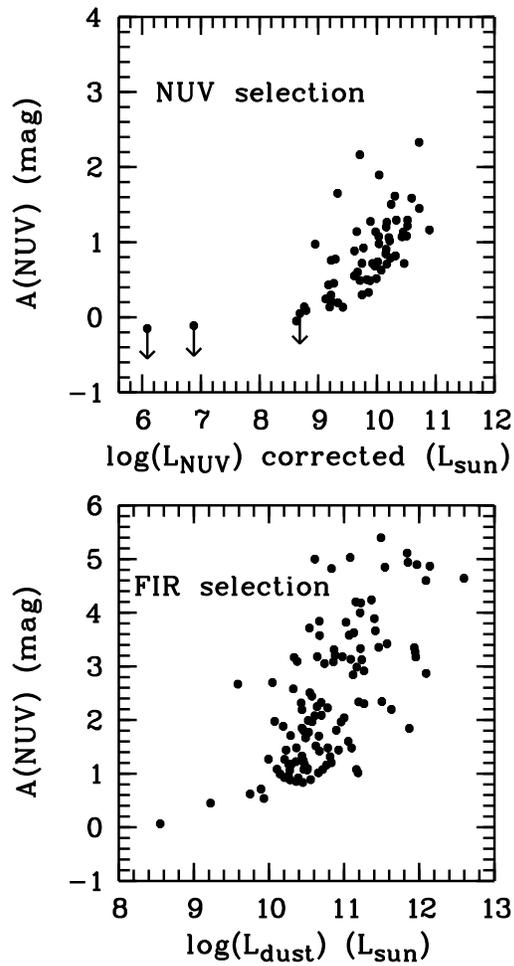}
   \caption{Dust attenuation in the GALEX NUV band against the luminosity of
the galaxies. Left panel: NUV selected sample, the luminosity is the NUV one
corrected for dust attenuation. Right panel: FIR selected sample, the
luminosity is the total dust luminosity}
      \label{fig3}
   \end{figure}
   \begin{figure}
\includegraphics{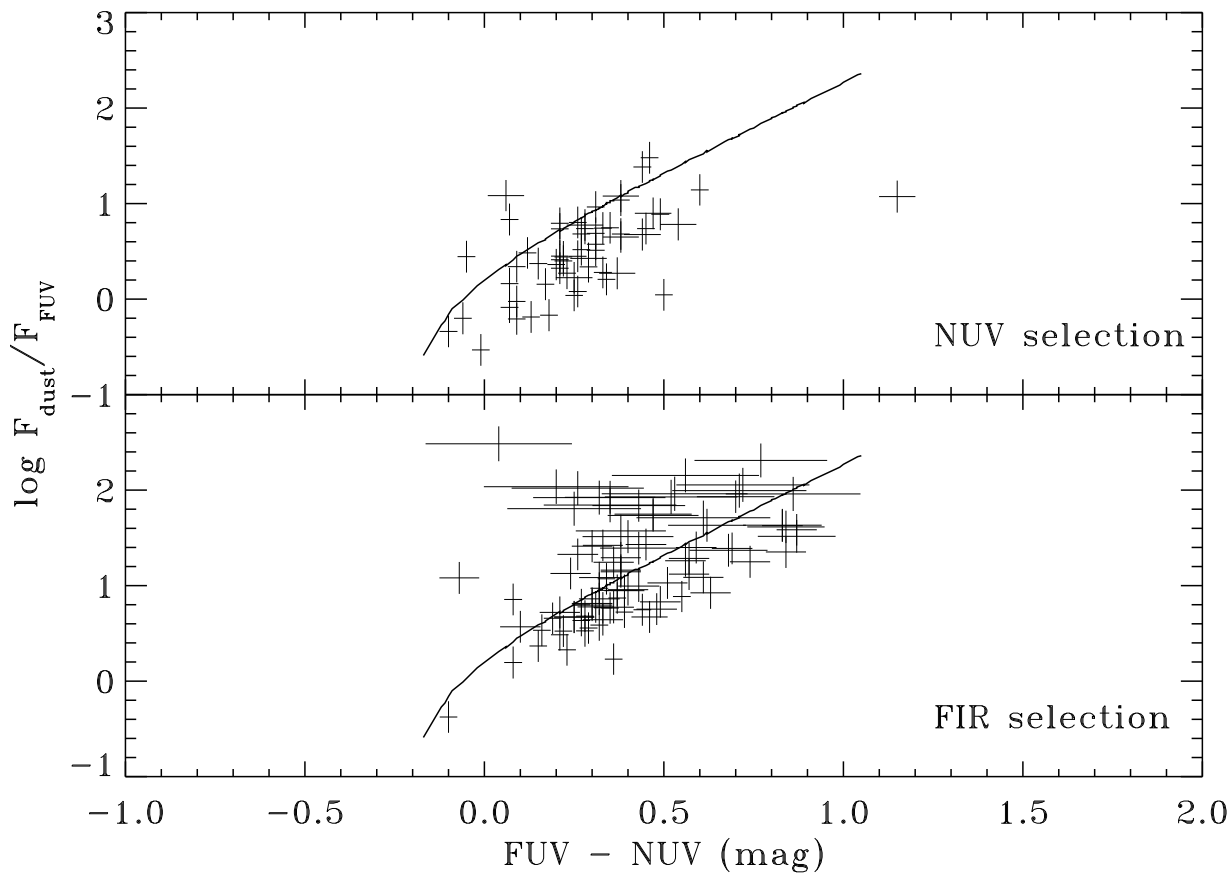}
   \caption{log($F_{\rm dust}/F_{\rm FUV}$) against the FUV-NUV color for NUV
and FIR
selected samples. The solid line is the mean relation expected for starburst
galaxies \citep{kon04}. The error bars were estimated assuming ~15\%
uncertainty for the
IRAS fluxes and ~40\% uncertainty for the $<F_{\rm dust}/F_{\rm FIR}>$ ratio,
the errors on the 
FUV and NUV fluxes were supposed to be poissonian.
}
      \label{fig4}
   \end{figure}

\end{document}